\documentclass[prd,preprintnumbers,superscriptaddress,nofootinbib,floatfix,twocolumn,notitlepage]{revtex4-2}
\usepackage[dvips]{graphicx}
\usepackage{dcolumn} 
\usepackage{bm}
\usepackage{epsfig,amsmath,amssymb,verbatim,mathrsfs,array,layout,textcomp,amssymb,latexsym,slashed}
\usepackage{color}
\usepackage[colorlinks=true,citecolor=blue,urlcolor=blue,linktocpage=true,
linkcolor=blue]{hyperref}
\usepackage[utf8]{inputenc}
\usepackage{multirow}

\input epsf.tex
\newcommand{\beq}{\begin{eqnarray}}
\newcommand{\eeq}{\end{eqnarray}}

\def\ltap{\ \raise.3ex\hbox{$<$\kern-.75em\lower1ex\hbox{$\sim$}}\ }
\def\gtap{\ \raise.3ex\hbox{$>$\kern-.75em\lower1ex\hbox{$\sim$}}\ }
\def\CO{{\cal O}}

\def\CO{{\cal O}}

\def\be{\begin{equation}}
\def\ee{\end{equation}}
\def\bea{\begin{eqnarray}}
\def\eea{\end{eqnarray}}

\pdfoutput=1

\def\mysection#1{{{\bf #1}.~}}

\newcommand{\Rinf}{R_\infty}

\definecolor{newred}{rgb}{0.5,0.1,0}
\definecolor{darkgreen}{rgb}{0.0,0.7,0.2}
\definecolor{lightblue}{rgb}{0.0,0.5,1} 

\RequirePackage[normalem]{ulem} 
\RequirePackage{color}\definecolor{RED}{rgb}{1,0,0}\definecolor{BLUE}{rgb}{0,0,1} 

\begin{document} 

\title{Towards an independent determination of muon g-2 from muonium spectroscopy}
\preprint{LAPTH-021/21}

\author{C\'edric Delaunay}
\email{cedric.delaunay@lapth.cnrs.fr}
\affiliation{Laboratoire d'Annecy-le-Vieux de Physique Th\'eorique LAPTh, CNRS -- USMB, BP 110 Annecy-le-Vieux, F-74941 Annecy, France}

\author{Ben Ohayon}
\email{bohayon@ethz.ch}
\affiliation{Institute for Particle Physics and Astrophysics, ETH Zürich, CH-8093 Zürich, Switzerland}

\author{Yotam Soreq}
\email{soreqy@physics.technion.ac.il}
\affiliation{Physics Department, Technion -- Israel Institute of Technology, Haifa 3200003, Israel}

\begin{abstract}
We show that  muonium spectroscopy in the coming years can reach a precision high enough to determine the anomalous magnetic moment of the muon below one part per million~(ppm). 
Such an independent determination of muon $g-2$, which is not limited by hadronic uncertainties, would certainly shed light on the $\sim2\,$ppm difference currently observed between spin-precession measurements and ($R$-ratio based) Standard Model predictions.
The magnetic dipole interaction between electrons and (anti)muons bound in muonium gives rise to a hyperfine splitting~(HFS) of the ground state which is sensitive to the muon anomalous magnetic moment. 
A direct comparison of the muonium frequency measurements of the HFS at J-PARC and the 1S-2S transition at PSI with theory predictions will allow to extract muon $g-2$ with high precision.
Improving the accuracy of QED calculations of these transitions by about one order of magnitude is also required.  
Moreover, the good agreement between theory and experiment for the electron $g-2$ indicates that new physics interactions are unlikely to affect muonium spectroscopy down to the envisaged precision.  
\end{abstract}

\maketitle

\mysection{Introduction}
The long and winding road that leads to the discovery of physics beyond the Standard Model~(SM) may well go through loops. 
The magnetic moments of the electron and the muon 
\beq
    \vec{\mu}_\ell
=   g_\ell \left(\frac{q_\ell}{2m_\ell}\right)\vec{S}_\ell \, , 
\eeq
($\ell=e,\mu$) are shining examples. 
Indeed, quantum fluctuations in the vacuum of all known {\it and unknown} fields inevitably causes their $g_\ell$-factors to deviate from the prediction of the Dirac equation by an ``anomalous'' part $a_\ell \equiv (g_\ell-2)/2$~\cite{Schwinger:1948iu}. 
Therefore, any accurate measurement of the latter, together with equally accurate theoretical predictions, open the door to new physics~(NP) phenomena.  

The Muon g-2 collaboration at the Fermi National Accelerator Laboratory~(FNAL) recently determined~\cite{Abi:2021gix,Albahri:2021kmg,Albahri:2021ixb} the muon anomalous magnetic moment to 460 parts-per-billion~(ppb) from spin-precession measurements. 
The central value is in good agreement with an earlier measurement~\cite{Bennett:2006fi} at the Brookhaven National Laboratory~(BNL), leading to a new experimental world average~\cite{Abi:2021gix}
\beq
    \label{eq:amuEXP}
    a_\mu^{\rm exp} 
=   116\,592\,061(41)\times 10^{-11} \, ,
\eeq
to $0.35$ parts-per-million~(ppm). 

The SM prediction for $a_\mu$ includes contributions from electromagnetic~\cite{Aoyama:2012wk,Aoyama:2019ryr,Czarnecki:2002nt,Gnendiger:2013pva,Davier:2017zfy,Keshavarzi:2018mgv,Colangelo:2018mtw,Hoferichter:2019gzf,Davier:2019can,Keshavarzi:2019abf,Kurz:2014wya,Melnikov:2003xd,Masjuan:2017tvw,Colangelo:2017fiz,Hoferichter:2018kwz,Gerardin:2019vio,Bijnens:2019ghy,Colangelo:2019uex,Blum:2019ugy,Colangelo:2014qya}, strong and weak interactions involving virtual exchange of photons, leptons, hadrons and $W$, $Z$ and Higgs bosons. 
Its evaluation was recently reviewed  by the international theory community, finding~\cite{Aoyama:2020ynm}
\beq
    \label{eq:amuSM}
    a_\mu^{\rm SM}
=   116\,591\,810(43) \times 10^{-11} \, ,
\eeq
where the theory uncertainty is dominated by uncertainties in the (nonperturbative) hadronic vacuum polarisation~(HVP) and hadronic light-by-light~(HLbL) contributions. 
This result is 4.2 standard deviation smaller than the experimental average, suggesting a possible contribution from NP with a magnitude of
\beq
    \label{eq:deltaamu}
    \Delta a_\mu
    \equiv 
    a_\mu^{\rm exp}-a_\mu^{\rm SM}
=   (251\pm 59)\times 10^{-11}\,,
\eeq
which is about $2\,$ppm. 
The SM prediction in Eq.~\eqref{eq:amuSM} relies on a leading-order~(LO) HVP contribution of $a_\mu^{\rm HVP-LO}(e^+e^-)=6\,931(40)\times 10^{-11}$~\cite{Aoyama:2020ynm,Davier:2019can,Davier:2017zfy,Keshavarzi:2018mgv,Colangelo:2018mtw,Hoferichter:2019gzf,Keshavarzi:2019abf} derived from experimental $e^+e^-\to$ hadrons data (the so-called $R$-ratio method) at low-energy using dispersion relations~\cite{Bouchiat:1961lbg,Brodsky:1967sr,Gourdin:1969dm}. 

An alternative determination of the LO-HVP contribution stems from an {\it ab-initio} calculation in lattice QCD~\cite{Blum:2002ii}. 
The latest result from the Budapest-Marseille-Wuppertal collaboration~(BMWc) $a_\mu^{\rm HVP-LO}({\rm lattice})=7\,075(55)\times 10^{-11}$~\cite{Borsanyi:2020mff}, consistent with previous calculations~\cite{Gerardin:2019rua,Davies:2019efs,Giusti:2019xct,Blum:2018mom,Borsanyi:2017zdw} but with a three-fold reduced uncertainty, is about $1\,$ppm larger than the $R$-ratio value and agrees with the experimental average in Eq.~\eqref{eq:amuEXP} within two standard deviations.\\

The present status of muon $g-2$ is therefore puzzling. 
Imagining that the discrepancy holds either between experiment and a converged theory (currently $2\,$ppm), or between the two theoretical calculations (currently $1\,$ppm), then an independent determination of $a_\mu$ to this level, demanding as it would be, is of immense importance. 

An alternative strategy for determining $a_\mu$ would be probing magnetic properties of muons bound in atoms instead of free particles. 
Muonium ($M\equiv \mu^+e^-$) is the bound state of an antimuon and an electron. 
It is a purely leptonic bound state, free of nuclear structure effects usually hampering the theoretical accuracy in ordinary atoms, for which high-order QED calculations are available. 
In contrast, muonium spectroscopy is currently limited by experimental uncertainties, being mostly of statistical origin~\cite{Jungmann:2016gak}. 
In this letter, we show that by pushing muonium spectroscopy to its limits, both theoretical and experimental, a determination of $a_\mu$ is possible with $\CO$($1\,$ppm) precision. This completely different approach would shed a new light on the current puzzle.\\

\mysection{Muon $g-2$ from muonium spectroscopy}
One of the most precisely measured spectral line in muonium is the hyperfine splitting~(HFS) of its 1S ground state $\nu_{\rm HFS}$ at $12\,$ppb~\cite{Liu:1999iz}. 
The HFS originates from the interaction of the electron spin with the magnetic field sourced by the magnetic dipole moment of the (static) antimuon. 
For electronic states without orbital momentum, the LO Hamiltonian is a Fermi contact interaction between the electron and muon magnetic moments~\cite{Cohen-Tannoudji:101367} 
\beq
    \label{eq:HHFS}
    H_{\rm HFS}
=   -\frac{2\mu_0}{3}\vec{\mu}_{e}\cdot\vec{\mu}_{\mu}\,\delta^3(r)\,, 
\eeq
where $\mu_0=2\alpha h/(e^2c)$ is the vacuum permeability. 
As it involves explicitly the magnetic moment of the muon, it is in principle sensitive to its anomalous part.

\begin{table}[tb]
\centering
   \begin{tabular}{c|c|ccc|c}
   \hline \hline
    $\nu_i$ & quantity &  & $u_r$  & & Refs.\\
    (unit) &  & current & ongoing & ultimate & \\
    \hline
    & QED        & 8.1& 5.7  & 0.7 & \cite{Meyer:1999cx,Karshenboim:2019iuq}\\
    & HVP & $\mathcal{O}(10^{-2})$ & & &\cite{Friar:1998wu} \\ 
    1S-2S &  $R_\infty$ &  1.9 & 0.65      & & \cite{2018-CODATA,Karshenboim:2019iuq}\\
    (ppt) &  $\alpha$ & $\mathcal{O}(10^{-3})$ &  & & \cite{2018-CODATA}\\

    & exp        & $3.99\times 10^3$   & 4.1  & 1.6 & \cite{Meyer:1999cx,Crivelli:2018vfe}\\
    \hline
    &  QED        &  16 & 2.2 & 0.2 & \cite{eides2016hyperfine,eidesOsaka2018} \\
    &  HVP        &  0.33 &0.18          &  &  \cite{2018-CODATA,Keshavarzi:2019abf} \\
    HFS &  $\alpha$ &  0.30 & 0.16   &   & \cite{2018-CODATA,Morel:2020dww}\\
    (ppb)&   $R_\infty$ & $\mathcal{O}(10^{-3})$ & & & \cite{2018-CODATA}\\
    &  exp       &  12     & 2.2 & 0.90 &\cite{Liu:1999iz,tanaka2021development}  \\
  \hline\hline
\end{tabular}
\caption{
Uncertainty budget for the 1S-2S transition (in ppt) and 1S HFS (in ppb) in muonium. 
The first column summarizes the current status of the relative uncertainties $u_r(x)=u(x)/x$. 
The second one (ongoing) indicates the milestone set by the Mu-MASS and MuSEUM experiments. 
The last one presents the foreseeable improvements necessary for a muonium determination of $a_\mu$ at sub-ppm level. 
Blank entries correspond to quantities that do not need further improvement.
} 
\label{tab:observables}
\end{table}

\begin{table}[tb]
    \centering
    \begin{tabular}{c|c|ccc}
\hline\hline
parameter & quantity &  & $u_r$  \\
 (unit)&  & current & ongoing & ultimate \\
\hline
            & $\nu_{\rm 1S-2S}$(exp) & 825 & 0.84 &  0.34 \\
$m_e/m_\mu$ & QED(1S-2S)      & 1.7 & 1.2 &  0.1  \\
   (ppb)         & $R_\infty$      & 0.40 & 0.13 &    \\
            & total           &  825& 1.5 &  0.37  \\  
\hline
         & $\nu_{\rm 1S-2S}$(exp) & 708 & 0.73 & 0.29  \\
         & $\nu_{\rm HFS}$(exp)   &  10 & 1.9 & 0.77  \\
         & QED(1S-2S)      & 1.4 & 1.0 & 0.07 \\
$a_\mu$  & QED(HFS)        & 14  & 1.9 & 0.2  \\
  (ppm)       & HVP(HFS)        & 0.29 & 0.16 &   \\
         & $R_\infty$      & 0.35 & 0.13 &  \\
         & $\alpha$        & 0.26 & 0.14 &   \\
         & total           & 708 & 3.0 & 0.88  \\     
\hline\hline         
    \end{tabular}
    \caption{Uncertainty budget for the determination of $m_e/m_\mu$ and $a_\mu$ from precision muonium spectroscopy. Same as Table~\ref{tab:observables}.}
    \label{tab:parameters}
\end{table}

Besides the fine-structure constant, $\alpha$, and the Rydberg constant $\Rinf=\alpha^2 m_ec/(2h)$, two parameters are endemic to muonic physics at low energy: the electron-muon mass ratio, $m_e/m_\mu$, and the muon magnetic moment.
Hydrogen spectroscopy determines $\Rinf$ at $1.9\,$ parts-per-trillion (ppt)~\cite{2018-CODATA}, making it the most accurately known constant in physics.\footnote{A more recent 1S-3S frequency measurement~\cite{Grinin1061} provides another determination of $R_\infty$ at $3.5\,$ppt consistent with the CODATA value.} 
Combining $\Rinf$ with accurate measurements of the electron and rubidium atomic masses and the $h/m_{\rm Rb}$ ratio gives the currently best determination of $\alpha$ at $81\,$ppt~\cite{Morel:2020dww}.\footnote{This value disagrees with the previous best determination of $\alpha$~\cite{Parker:2018vye} 
by $5.4\sigma$~\cite{Morel:2020dww}, possibly due to systematic errors~\cite{2020Natur.588...37M}. 
While puzzling, this discrepancy is too small (relative to uncertainties in muonium) to affect significantly present and future determinations of $a_\mu$.} 
The muonic constants must be extracted from (at least) two other independent observables.
$m_e/m_\mu$ is known at $19\,$ppb from comparing the measured ground-state HFS of muonium~\cite{Liu:1999iz} with the SM prediction~\cite{Eides:2018rph}. 
Since we envisage to use this observable to extract $a_\mu$, $m_e/m_\mu$ must be obtained by other means.
The current second best determination of the electron-muon mass ratio is at $120\,$ppb, coming from a measurement of the (total) muon magnetic moment~\cite{Liu:1999iz}. 
However, this cannot be used either as it clearly depends on $a_\mu$. 
Another way to extract $m_e/m_\mu$ is to measure a muonium line that is (mostly) independent of the magnetic moment. 
To date, the only possibility is the Lyman-$\alpha$ line between the 1S and 2S states.  \\

The theory prediction for the 1S-2S transition frequency in muonium is
\beq
    \label{eq:M1S2Sth}
    \nu_{\rm 1S-2S} 
=   \frac{3}{4}\frac{\Rinf c}{(1+m_e/m_\mu)}
    \left[1+\delta_{\rm 1S-2S}\right]\,,
\eeq
where the muon mass enters as a recoil contribution of $\sim0.5\%$ and subleading corrections in $\delta_{\rm 1S-2S}\sim\CO(\alpha^2)$ are known up to three-loop QED with 20$\,$kHz uncertainty~\cite{Meyer:1999cx}. 

For the ground-state HFS the theory prediction reads
\beq
    \label{eq:MHFSth}
    \nu_{\rm HFS}
=   \frac{16}{3}(1+a_\mu)\frac{m_e}{m_\mu}   
    \frac{\Rinf c \alpha^2}{(1+m_e/m_\mu)^3}
    \left[1+\delta_{\rm HFS}\right]\,,
\eeq
where $\delta_{\rm HFS}\sim\CO(\alpha/\pi)$ gathers corrections~\cite{2016-+CODATA14} beyond the Hamiltonian in Eq.~\eqref{eq:HHFS} from relativistic, radiative (including the anomalous magnetic moment $a_e$ of the electron), recoil, radiative-recoil, weak and hadronic contributions. 
The theory uncertainty $\delta_{\rm HFS}$ is about 70$\,$Hz~\cite{eides2016hyperfine,Eides:2018rph}, dominated by unknown three-loop QED contributions to the radiative-recoil term in $\delta_{\rm HFS}$~\cite{Eides:2018rph}.

Since the $\delta_{\rm HFS,1S-2S}$ corrections above only weakly depend on fundamental constants, the leading  contributions to $\nu_{\rm HFS}$ and $\nu_{\rm 1S-2S}$ are sufficient to estimate the sensitivity of muonium spectroscopy to muon $g-2$. 
Combining Eqs.~\eqref{eq:M1S2Sth} and~\eqref{eq:MHFSth} allows to separately determine $m_e/m_\mu$ and $a_\mu$, providing a simple estimate of their uncertainties
\beq
    \label{eq:umemmu}
    u^2(m_e/m_\mu)
    &\simeq& 
    u_r^2(\Rinf)+u_r^2(\nu_{\rm 1S-2S})\nonumber\\
    &&+\delta_{\rm 1S-2S}^2 u_r^2(\delta_\text{1S-2S})\,,\\
    \label{eq:uamu}
    u^2(a_\mu)
    &\simeq&
    u_r^2(m_e/m_\mu)+4u_r^2(\alpha)+u_r^2(\nu_{\rm HFS})\nonumber\\
    &&+\delta_{\rm HFS}^2 u_r^2(\delta_\text{HFS})\,,
\eeq
where $u_r(x)\equiv u(x)/x$ and $u(x)$ is one standard deviation of the observable $x$, and $u_r(\delta_i)$ denotes the relative theory uncertainty of $\delta_i$ that is not associated with the $\Rinf$, $\alpha$ and $m_e/m_\mu$ parameters. 

A least-square adjustment (see Supplemental Material, which includes Refs. \cite{Mohr:2000ie,Mariam:1982bq,Karshenboim:2021jsc,Eides:2014xea}) using 2018 CODATA recommended values for $\Rinf$ and $\alpha$, and including the state-of-the-art calculation (see Ref.~\cite{2018-CODATA} and references therein) of $\delta_{\rm HFS, 1S-2S}$ yields
\beq
    	\label{eq:amuM}
    	a_\mu^{M}
=   	116\,637(82)\times 10^{-8}\,,
\eeq
and $m_e/m_\mu=4\,836\,329(4)\times 10^{-9}$, which is larger ($a_\mu^M-a_\mu^{\rm exp}\simeq 4.5\times 10^{-7}$) but consistent with both the experimental value in Eq.~\eqref{eq:amuEXP} and the theoretical result in Eq.~\eqref{eq:amuSM}.
The large uncertainty $u(a_\mu)=8.2\times 10^{-7}$ dominated by the experimental $u_r(\nu_{\rm 1S-2S})=4.0\times 10^{-9}$~\cite{Meyer:1999cx}. 
Note that this determination of $a_\mu$ assumes that muonium theory follows SM predictions. 
As shown below, contributions beyond the SM ones related to an hypothetical NP coupling to electrons are sufficiently constrained not to affect the least-square adjustment.\\

\mysection{Expected improvements in muonium physics}
The present data is lacking precision to provide a competitive determination of muon $g-2$. 
However the situation is expected to dramatically improve in the near future thanks to new experimental techniques and more accurate QED calculations. 
We summarize 
the developments planned at the next round of experiments at PSI and J-PARC and show that, together with the ongoing theory improvement, they will allow for an extraction of $a_\mu$ at few ppm. 
Moreover, we outline and argue on the feasibility of the refinements necessary to bring $u_r(a_\mu)$ below the ppm-level.\\    

\textit{1S-2S transition}
The first necessary ingredient is to improve the electron-muon mass ratio from muonium 1S-2S spectroscopy. 
Spectroscopy of the 1S-2S transition in hydrogenic atoms relies on two-photon excitation with a UV laser, operating at a $244\,$nm wavelength for muonium. 
The very high transition frequency makes possible measurements at ppt precision, with a $145$\,kHz natural linewidth due to the muon lifetime.

One of the main challenges is the low excitation efficiency, as the 1S-2S transition is a two-photon transition.
In order to increase the transition probability, previous 1S-2S measurements utilized a high-power pulsed laser. 
This however came at the cost of broadening the linewidth to $20\,$MHz associated with the laser pulse-width. 
Another main systematic uncertainty of $10$\,MHz also originated from the pulsed interaction. 
A high-power pulsed laser changes its frequency during every pulse, an effect known as chirping which is notoriously difficult to compensate for.

To circumvent the limitations of pulsed laser excitation, the MuoniuM lAser SpectroScopy~(Mu-MASS) experiment~\cite{Crivelli:2018vfe} at PSI utilizes a cavity-enhanced continuous wave~(CW) excitation~\cite{burkley2021stable}. 
The reduced excitation efficiency in CW operation is compensated by the use of the low-energy-muon~(LEM) beamline~\cite{PhysRevLett.72.2793} paired with new methods to obtain slow muonium atoms emitted into vacuum after production in mesoporous thin SiO$_2$ films~\cite{PhysRevLett.108.143401}. 
With such techniques the Mu-MASS uncertainty goal was set to $10\,$kHz ($4\,$ppt)~\cite{Crivelli:2018vfe}. 

On the theory side the $\nu_{\rm 1S-2S}$ uncertainty was quoted at $20\,$kHz in Ref.~\cite{Meyer:1999cx}. 
To the best of our knowledge this figure has not been updated, despite the recent improvement in QED calculations for hydrogen-like atoms~\cite{Yerokhin:2018gna,Karshenboim:2019iuq}. 
By rescaling the theory uncertainties for hydrogen~\cite{2018-CODATA} to the muon mass (discarding nuclear finite-size and polarizability contributions) we assess the current theory uncertainty of $\nu_{\rm 1S-2S}$ to be at $14\,$kHz ($5.7\,$ppt). 

The Rydberg constant is also expected to improve in the next few years, anticipating a full resolution of the so-called proton radius puzzle~\cite{karr2020proton}. 
The very precise determination of the proton radius from muonic hydrogen~\cite{Pohl:2010zza,Antognini:1900ns} improves the theoretical precision in hydrogen by about one order-of-magnitude, which then becomes limited by bound-state QED calculations. 
The current QED uncertainty in hydrogen is roughly $1\,$kHz~\cite{Karshenboim:2019iuq}, allowing in principle a three-fold~\cite{PhysRevA.93.022513} more precise determination of $\Rinf$ relative to the latest CODATA~\cite{2018-CODATA}. 

Using Eq.~\eqref{eq:umemmu} the above values (referred to as ``ongoing" in Tables~\ref{tab:observables} and~\ref{tab:parameters}) yield an expected precision on $m_e/m_\mu$ at $1.5\,$ppb.\\

\textit{Ground state HFS}
The major improvement to the electron-muon mass ratio considered above opens up the possibility to obtain a value for $a_\mu$ with few ppm uncertainty, comparable to the current difference in Eq.~\eqref{eq:deltaamu}, granted that the ground state HFS is improved as well. 
The current best measurement of $\nu_{\rm HFS}$ was done with a chopped beam at LAMPF and limited by statistics~\cite{Liu:1999iz}.
The Muonium Spectroscopy Experiment Using Microwave~(MuSEUM) experiment will improve the statistical uncertainty by using the high-intensity pulsed muon beam at J-PARC~\cite{torii2015precise,Ueno:2018hdr} as well as Rabi-oscillation spectroscopy~\cite{nishimura2021rabioscillation}. 
Moreover, recently a thorough optimization of the microwave cavity has been done to drive down the systematic uncertainties to the ppb level~\cite{tanaka2021development}. 
This will allow for more precise HFS measurements by about one order-of-magnitude~\cite{tanaka2021development} compared to previous LAMPF measurements.
We take $10\,$Hz ($2.2\,$ppb) as an estimate for the uncertainty goal of the MuSEUM experiment, which is compatible with the `several ppb' in.~\cite{tanaka2021development}. (See also Refs.~\cite{asaka2018precision,1868388} and references therein.)  
Another important systematic uncertainty is due to the pressure shift from the finite gas density environment~\cite{Kanda:2020mmc} in which HFS measurement are performed.
This can be overcome by measuring the HFS in vacuum using a low-energy muon beam.

Excluding the muon mass uncertainty, the theoretical ground state HFS calculation is currently limited by bound-state QED to around $70\,$Hz accuracy~\cite{Eides:2018rph} coming from uncalculated sets of three-loop diagrams.
Efforts to improve this calculation are ongoing, quoting a goal of $10\,$Hz~\cite{eides2016hyperfine}.

Collecting the theoretical and experimental values discussed above we estimate with Eq.~\eqref{eq:uamu} the expected uncertainty for extracting $a_\mu$ from the ongoing effort in muonium spectroscopy as about $3\,$ppm (shown in blue on Fig.~\ref{fig:amu}). 
Such rousing prospects, comparable to the difference in Eq.~\eqref{eq:deltaamu}, would already contribute to the muon $g-2$ puzzle.\\

\textit{Further improvements} 
Next, we explore how much the above uncertainties can be further reduced. 
We based our estimations on known experimental techniques, arguing that the necessary theoretical improvements can be reasonably achieved along the way.

An improvement on the 1S-2S frequency determination from the Mu-MASS goal of $10\,$kHz to few kHz is expected~\cite{Crivelli:2018vfe}. 
Since final systematic uncertainties are estimated at the kHz-level~\cite{Crivelli:2018vfe}, this precision could be accommodated by an order-of-magnitude increase in statistics. 
Such an increase could either come from an improved high-energy muon beam rate, as considered in the High-intensity Muon Beam~(HIMB) upgrade~\cite{kiselev2021progress, kiselev2021status, knecht2017high} at PSI, and/or a higher efficiency in muon moderation, as is pursued by the MuCool collaboration~\cite{belosevic2019mucool,PhysRevLett.125.164802,antogninimucool}. 
Moreover, another effort in improving the 1S-2S measurement is under consideration at J-PARC~\cite{Keshavarzi:2021eqa}.
A 1S-2S measurement at $4\,$kHz would be sufficient to make the $m_e/m_\mu$ uncertainty on $\nu_{\rm HFS}$ subdominant. 

Such an improved experimental precision must be supplemented by higher accuracy calculations.
Following recent progress in such calculations~\cite{Yerokhin:2018gna, Karshenboim:2019iuq,Eides:2021wuv}, the main limitation for the theoretical uncertainty comes from unknown radiative-recoil corrections of  $\mathcal{O}[\alpha (Z\alpha)^6 (m/M)]$~\cite{Karshenboim:2019iuq}.  This correction is likely to be calculated in the near future since the recent convergence on the proton radius~\cite{grinin2020two, karr2020proton, Ubachs1033,Bezginov1007, HAMMER2020257, cui2021fresh} makes it the limiting factor to reducing the Rydberg constant uncertainty, as well as the Deuterium radius deduced from the H-D isotope shift~\cite{Pachucki:2018yxe}. 
The pure recoil correction at $\mathcal{O}[(Z\alpha)^6 (m/M)^2]$ is also required for kHz-accuracy in M. It is partially calculated~\cite{Blokland:2001fn} and there is no known obstacle  towards a complete result.\\

Regarding the ground state HFS, a further improvement upon the ongoing efforts, both in experiment and theory, is more demanding. 
Experimentally the muon-lifetime constraint on the linewidth already poses a major challenge. 
A measurement at $10\,$Hz precision already requires to resolve the line center to $~10^{-4}$ of the linewidth, an achievement similar to the recent 2S-4P frequency measurement in hydrogen~\cite{Beyer:2017gug}. 
Further improving the precision with better line-shape modeling would constitute a premiere in spectroscopy. 
A complementary approach consists in converting extremely high statistics into narrower linewidths by post-selecting so-called `old muonium' atoms~\cite{Liu:1995hq,Liu:1999iz}, namely muons which have not decayed after several lifetimes. 
Therefore, despite the challenges, we imagine that HFS measurements could be done at the ppb-level ($\simeq4\,$Hz). 

As regards systematic uncertainty, down to the ppb-level the main one is still considered to be the quadratic pressure shift~\cite{Liu:1999iz}, which could be reduced with lower pressures, relying on higher statistics. 
Another promising option would be to mix gases such as He and Kr which have opposite pressure shift contributions~\cite{PhysRevA.2.1411}.
The needed high rates could be accommodated by the HIMB upgrade at PSI that will deliver $\sim 10^{10}$ antimuons per second, which is three orders of magnitude higher than the CW muon beam used for the HFS measurement at LAMPF.
Leveraging the technical advancements by the MuSEUM collaboration~\cite{nishimura2021rabioscillation,tanaka2021development}, together with a  $\sim100$-days beamtime, we conclude that a sub-ppb experimental uncertainty in $\nu_{\rm HFS}$ could be envisioned.

The corresponding theory prediction to the level of $4$\,Hz accuracy is demanding as well. 
However, as long as the uncertainty from HVP contributions (presently about $0.8\,$Hz~\cite{Keshavarzi:2019abf}) is subdominant, completing the required bound-state QED calculations in the same timescale as experimental milestones could be reasonably envisaged. 
Indeed, ongoing work already set a goal of a few Hz~\cite{eidesOsaka2018}, which would suffice to make $\nu_{\rm HFS}$ limited by experiment. \\

\begin{figure}
    \centering
    \includegraphics[width=\columnwidth]{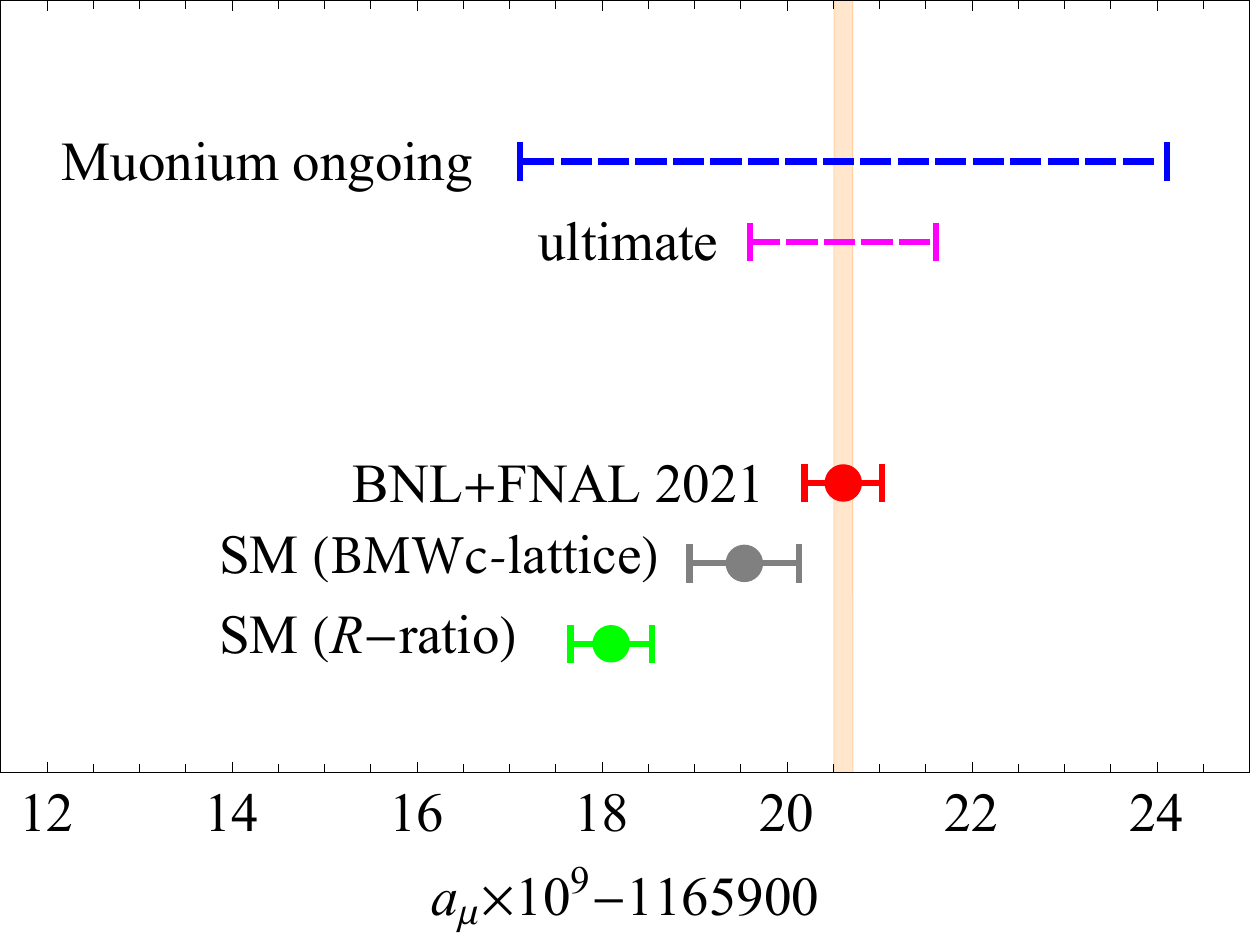}
    \caption{Anomalous magnetic moment of the muon determined from spin-precession measurements at BNL and FNAL (red)~\cite{Abi:2021gix}, Standard Model calculations with LO HVP evaluated from $e^+e^-\to$ hadrons data (green)~\cite{Aoyama:2020ynm} or lattice QCD (gray)~\cite{Borsanyi:2020mff} and the projected sensitivities in muonium (dashed) with the currently planned spectroscopy improvements (blue) and its ultimate improvement (magenta), centered around the current experimental average. The orange band shows the four-fold improved FNAL standard deviation expected in the near future.
  }
    \label{fig:amu}
\end{figure}

Assuming that the 1S-2S transition and ground state HFS in muonium are pushed to the aforementioned limits would bring down the $a_\mu$ uncertainty to $u(a_\mu)\simeq 103\times 10^{-11}$ ($0.88$~ppm) which is about half of the current difference between theory and experiment.\\

\mysection{Summary and Outlook}
We proposed an alternative strategy to extract the anomalous magnetic moment of the muon based on highly-precise measurements and theoretical QED calculations in muonium. 
Current data on the 1S-2S transition and ground state HFS frequencies allows to determine $a_\mu$ at $\sim700\,$ppm, largely limited by statistics in the 1S-2S measurement. 
A new program, partly motivated by the present muon $g-2$ puzzle, of highly-precise muonium spectroscopy and theory will significantly improve the situation in the coming decade, reducing the $a_\mu$ uncertainty to about $3\,$ppm, and even below the ppm-level at a future stage. 
As shown in the Supplemental Material, which includes Refs. \cite{Frugiuele:2019drl, Karshenboim:2010cg, Jackiw:1972jz, Jegerlehner:2009ry, Balkin:2021rvh, messiah_2014, Salpeter:1951sz, Hanneke:2008tm, Hardy:2016kme, Fan:2020ijg,Gabrielse:2019cgf, Ilten:2018crw,Andreev:2021xpu,CODATANP}, this indirect determination which relies on bound-state QED calculation is unlikely to be affected by NP, given present constraints on exotic forces coupled to electrons.\\ 

Such an independent value of $a_\mu$ cannot be competitive with spin-precession measurements that are expected to reach $\sim0.1\,$ppm accuracy in a few years~\cite{Abi:2021gix,Abe:2019thb}. 
However it would most certainly help interpreting the current difference between these measurements and the ($R$-ratio based) SM prediction, were it to persist over the same timescale. 
A muonium value of $a_\mu$ favoring today's experimental average, along with $\delta_{\rm 1S-2S}$ and $\delta_{\rm HFS}$ consistent with the SM, would strengthen the evidence of NP dominantly coupled to muons (assuming the $R$-ratio determination of the HVP contribution is a faithful figure of the SM prediction).  
Conversely, a value consistent with the current $R$-ratio prediction would indicate a potential unknown systematic with the BNL/FNAL measurements, or point to a non-trivial NP dynamics. In the latter case, precision measurements of additional muonium lines, like the 2S-2P Lamb shift~\cite{Janka:2020xky,ohayon2021precision} ongoing at PSI, would help discriminating among  different possibilities. 
Either way, this highly motivates to push the frontier of precision muonium physics as much as possible.\\

\mysection{Acknowledgements} 
We thank Michael Eides and Yevgeny Stadnik for useful discussions and 
Paolo Crivelli, Michael Eides, Gilad Perez and in particular Klaus Jungmann for valuable comments on the manuscript. 
BO acknowledges support from the European Union’s Horizon 2020 research and innovation program under the Marie Skłodowska-Curie grant agreement No.~101019414.
YS is supported by the United States-Israel Binational Science Foundation (BSF) (NSF-BSF program grant No.~2018683), by the Israel Science Foundation (grant No.~482/20) and by the Azrieli foundation. 
YS is Taub fellow (supported by the Taub Family Foundation).

\bibliographystyle{unsrt}
\bibliography{Mrefs}

\begin{thebibliography}{100}

\bibitem{Schwinger:1948iu}
Julian~S. Schwinger.
\newblock {On Quantum electrodynamics and the magnetic moment of the electron}.
\newblock {\em Phys. Rev.}, 73:416--417, 1948.

\bibitem{Abi:2021gix}
B.~Abi et~al.
\newblock {Measurement of the Positive Muon Anomalous Magnetic Moment to
  0.46~ppm}.
\newblock {\em Phys. Rev. Lett.}, 126(14):141801, 2021.

\bibitem{Albahri:2021kmg}
T.~Albahri et~al.
\newblock {Magnetic-field measurement and analysis for the Muon $g-2$
  Experiment at Fermilab}.
\newblock {\em Phys. Rev. A}, 103(4):042208, 2021.

\bibitem{Albahri:2021ixb}
T.~Albahri et~al.
\newblock {Measurement of the anomalous precession frequency of the muon in the
  Fermilab Muon $g-2$ Experiment}.
\newblock {\em Phys. Rev. D}, 103(7):072002, 2021.

\bibitem{Bennett:2006fi}
G.~W. Bennett et~al.
\newblock {Final Report of the Muon E821 Anomalous Magnetic Moment Measurement
  at BNL}.
\newblock {\em Phys. Rev. D}, 73:072003, 2006.

\bibitem{Aoyama:2012wk}
Tatsumi Aoyama, Masashi Hayakawa, Toichiro Kinoshita, and Makiko Nio.
\newblock {Complete Tenth-Order QED Contribution to the Muon $g-2$}.
\newblock {\em Phys. Rev. Lett.}, 109:111808, 2012.

\bibitem{Aoyama:2019ryr}
Tatsumi Aoyama, Toichiro Kinoshita, and Makiko Nio.
\newblock {Theory of the Anomalous Magnetic Moment of the Electron}.
\newblock {\em Atoms}, 7(1):28, 2019.

\bibitem{Czarnecki:2002nt}
Andrzej Czarnecki, William~J. Marciano, and Arkady Vainshtein.
\newblock {Refinements in electroweak contributions to the muon anomalous
  magnetic moment}.
\newblock {\em Phys. Rev.}, D67:073006, 2003.
\newblock [Erratum: Phys. Rev. {\bf D73}, 119901 (2006)].

\bibitem{Gnendiger:2013pva}
C.~Gnendiger, D.~St{\"o}ckinger, and H.~St{\"o}ckinger-Kim.
\newblock {The electroweak contributions to $(g-2)_\mu$ after the Higgs boson
  mass measurement}.
\newblock {\em Phys. Rev.}, D88:053005, 2013.

\bibitem{Davier:2017zfy}
Michel Davier, Andreas Hoecker, Bogdan Malaescu, and Zhiqing Zhang.
\newblock {Reevaluation of the hadronic vacuum polarisation contributions to
  the Standard Model predictions of the muon $g-2$ and ${\alpha (m_Z^2)}$ using
  newest hadronic cross-section data}.
\newblock {\em Eur. Phys. J.}, C77(12):827, 2017.

\bibitem{Keshavarzi:2018mgv}
Alexander Keshavarzi, Daisuke Nomura, and Thomas Teubner.
\newblock {Muon $g-2$ and $\alpha(M_Z^2)$: a new data-based analysis}.
\newblock {\em Phys. Rev.}, D97(11):114025, 2018.

\bibitem{Colangelo:2018mtw}
Gilberto Colangelo, Martin Hoferichter, and Peter Stoffer.
\newblock {Two-pion contribution to hadronic vacuum polarization}.
\newblock {\em JHEP}, 02:006, 2019.

\bibitem{Hoferichter:2019gzf}
Martin Hoferichter, Bai-Long Hoid, and Bastian Kubis.
\newblock {Three-pion contribution to hadronic vacuum polarization}.
\newblock {\em JHEP}, 08:137, 2019.

\bibitem{Davier:2019can}
M.~Davier, A.~Hoecker, B.~Malaescu, and Z.~Zhang.
\newblock {A new evaluation of the hadronic vacuum polarisation contributions
  to the muon anomalous magnetic moment and to
  $\mathbf{\boldsymbol\alpha(m_Z^2)}$}.
\newblock {\em Eur. Phys. J.}, C80(3):241, 2020.
\newblock [Erratum: Eur. Phys. J. {\bf C80}, 410 (2020)].

\bibitem{Keshavarzi:2019abf}
Alexander Keshavarzi, Daisuke Nomura, and Thomas Teubner.
\newblock {The $g-2$ of charged leptons, $\alpha(M_Z^2)$ and the hyperfine
  splitting of muonium}.
\newblock {\em Phys. Rev.}, D101:014029, 2020.

\bibitem{Kurz:2014wya}
Alexander Kurz, Tao Liu, Peter Marquard, and Matthias Steinhauser.
\newblock {Hadronic contribution to the muon anomalous magnetic moment to
  next-to-next-to-leading order}.
\newblock {\em Phys. Lett.}, B734:144--147, 2014.

\bibitem{Melnikov:2003xd}
Kirill Melnikov and Arkady Vainshtein.
\newblock {Hadronic light-by-light scattering contribution to the muon
  anomalous magnetic moment revisited}.
\newblock {\em Phys. Rev.}, D70:113006, 2004.

\bibitem{Masjuan:2017tvw}
Pere Masjuan and Pablo S{\'a}nchez-Puertas.
\newblock {Pseudoscalar-pole contribution to the $(g_{\mu}-2)$: a rational
  approach}.
\newblock {\em Phys. Rev.}, D95(5):054026, 2017.

\bibitem{Colangelo:2017fiz}
Gilberto Colangelo, Martin Hoferichter, Massimiliano Procura, and Peter
  Stoffer.
\newblock {Dispersion relation for hadronic light-by-light scattering: two-pion
  contributions}.
\newblock {\em JHEP}, 04:161, 2017.

\bibitem{Hoferichter:2018kwz}
Martin Hoferichter, Bai-Long Hoid, Bastian Kubis, Stefan Leupold, and
  Sebastian~P. Schneider.
\newblock {Dispersion relation for hadronic light-by-light scattering: pion
  pole}.
\newblock {\em JHEP}, 10:141, 2018.

\bibitem{Gerardin:2019vio}
Antoine G{\'e}rardin, Harvey~B. Meyer, and Andreas Nyffeler.
\newblock {Lattice calculation of the pion transition form factor with
  $N_f=2+1$ Wilson quarks}.
\newblock {\em Phys. Rev.}, D100(3):034520, 2019.

\bibitem{Bijnens:2019ghy}
Johan Bijnens, Nils Hermansson-Truedsson, and Antonio
  Rodr{\'i}guez-S{\'a}nchez.
\newblock {Short-distance constraints for the HLbL contribution to the muon
  anomalous magnetic moment}.
\newblock {\em Phys. Lett.}, B798:134994, 2019.

\bibitem{Colangelo:2019uex}
Gilberto Colangelo, Franziska Hagelstein, Martin Hoferichter, Laetitia Laub,
  and Peter Stoffer.
\newblock {Longitudinal short-distance constraints for the hadronic
  light-by-light contribution to $(g-2)_\mu$ with large-$N_c$ Regge models}.
\newblock {\em JHEP}, 03:101, 2020.

\bibitem{Blum:2019ugy}
Thomas Blum, Norman Christ, Masashi Hayakawa, Taku Izubuchi, Luchang Jin,
  Chulwoo Jung, and Christoph Lehner.
\newblock {The hadronic light-by-light scattering contribution to the muon
  anomalous magnetic moment from lattice QCD}.
\newblock {\em Phys. Rev. Lett.}, 124(13):132002, 2020.

\bibitem{Colangelo:2014qya}
Gilberto Colangelo, Martin Hoferichter, Andreas Nyffeler, Massimo Passera, and
  Peter Stoffer.
\newblock {Remarks on higher-order hadronic corrections to the muon $g-2$}.
\newblock {\em Phys. Lett.}, B735:90--91, 2014.

\bibitem{Aoyama:2020ynm}
T.~Aoyama et~al.
\newblock {The anomalous magnetic moment of the muon in the Standard Model}.
\newblock {\em Phys. Rept.}, 887:1--166, 2020.

\bibitem{Bouchiat:1961lbg}
Claude Bouchiat and Louis Michel.
\newblock {La r\'esonance dans la diffusion m\'eson
  \ensuremath{\pi}\textemdash{} m\'eson \ensuremath{\pi} et le moment
  magn\'etique anormal du m\'eson \ensuremath{\mu}}.
\newblock {\em J. Phys. Radium}, 22(2):121--121, 1961.

\bibitem{Brodsky:1967sr}
Stanley~J. Brodsky and Eduardo De~Rafael.
\newblock {SUGGESTED BOSON - LEPTON PAIR COUPLINGS AND THE ANOMALOUS MAGNETIC
  MOMENT OF THE MUON}.
\newblock {\em Phys. Rev.}, 168:1620--1622, 1968.

\bibitem{Gourdin:1969dm}
M.~Gourdin and E.~De~Rafael.
\newblock {Hadronic contributions to the muon g-factor}.
\newblock {\em Nucl. Phys. B}, 10:667--674, 1969.

\bibitem{Blum:2002ii}
T.~Blum.
\newblock {Lattice calculation of the lowest order hadronic contribution to the
  muon anomalous magnetic moment}.
\newblock {\em Phys. Rev. Lett.}, 91:052001, 2003.

\bibitem{Borsanyi:2020mff}
Sz. Borsanyi et~al.
\newblock {Leading hadronic contribution to the muon 2 magnetic moment from
  lattice QCD}.
\newblock {\em Nature}, 593(7857):51--55, 2021.

\bibitem{Gerardin:2019rua}
Antoine G\'erardin, Marco C\`e, Georg von Hippel, Ben H\"orz, Harvey~B. Meyer,
  Daniel Mohler, Konstantin Ottnad, Jonas Wilhelm, and Hartmut Wittig.
\newblock {The leading hadronic contribution to $(g-2)_\mu$ from lattice QCD
  with $N_{\rm f}=2+1$ flavours of O($a$) improved Wilson quarks}.
\newblock {\em Phys. Rev. D}, 100(1):014510, 2019.

\bibitem{Davies:2019efs}
C.~T.~H. Davies et~al.
\newblock {Hadronic-vacuum-polarization contribution to the
  muon\textquoteright{}s anomalous magnetic moment from four-flavor lattice
  QCD}.
\newblock {\em Phys. Rev. D}, 101(3):034512, 2020.

\bibitem{Giusti:2019xct}
D.~Giusti, V.~Lubicz, G.~Martinelli, F.~Sanfilippo, and S.~Simula.
\newblock {Electromagnetic and strong isospin-breaking corrections to the muon
  $g - 2$ from Lattice QCD+QED}.
\newblock {\em Phys. Rev. D}, 99(11):114502, 2019.

\bibitem{Blum:2018mom}
T.~Blum, P.~A. Boyle, V.~G\"ulpers, T.~Izubuchi, L.~Jin, C.~Jung, A.~J\"uttner,
  C.~Lehner, A.~Portelli, and J.~T. Tsang.
\newblock {Calculation of the hadronic vacuum polarization contribution to the
  muon anomalous magnetic moment}.
\newblock {\em Phys. Rev. Lett.}, 121(2):022003, 2018.

\bibitem{Borsanyi:2017zdw}
Sz. Borsanyi et~al.
\newblock {Hadronic vacuum polarization contribution to the anomalous magnetic
  moments of leptons from first principles}.
\newblock {\em Phys. Rev. Lett.}, 121(2):022002, 2018.

\bibitem{Jungmann:2016gak}
Klaus~P. Jungmann.
\newblock {Precision Muonium Spectroscopy}.
\newblock {\em J. Phys. Soc. Jap.}, 85(9):091004, 2016.

\bibitem{Liu:1999iz}
Weiwen Liu et~al.
\newblock {High precision measurements of the ground state hyperfine structure
  interval of muonium and of the muon magnetic moment}.
\newblock {\em Phys. Rev. Lett.}, 82:711--714, 1999.

\bibitem{Cohen-Tannoudji:101367}
Claude Cohen-Tannoudji, Bernard Diu, and Franck Laloë.
\newblock {\em {Quantum mechanics; 1st ed.}}
\newblock Wiley, New York, NY, 1977.
\newblock Trans. of : Mécanique quantique. Paris : Hermann, 1973.

\bibitem{Meyer:1999cx}
V.~Meyer et~al.
\newblock {Measurement of the 1s - 2s energy interval in muonium}.
\newblock {\em Phys. Rev. Lett.}, 84:1136, 2000.

\bibitem{Karshenboim:2019iuq}
Savely~G. Karshenboim, Akira Ozawa, Valery~A. Shelyuto, Robert Szafron, and
  Vladimir~G. Ivanov.
\newblock {The Lamb shift of the 1$s$ state in hydrogen: Two-loop and
  three-loop contributions}.
\newblock {\em Phys. Lett. B}, 795:432--437, 2019.

\bibitem{Friar:1998wu}
James~Lewis Friar, J.~Martorell, and D.~W.~L. Sprung.
\newblock {Hadronic vacuum polarization and the Lamb shift}.
\newblock {\em Phys. Rev. A}, 59:4061--4063, 1999.

\bibitem{2018-CODATA}
Eite Tiesinga, Peter~J Mohr, David~B Newell, and Barry~N Taylor.
\newblock The 2018 codata recommended values of the fundamental physical
  constants.
\newblock {\em Web Version}, 8, 2019.

\bibitem{Crivelli:2018vfe}
P.~Crivelli.
\newblock {The Mu-MASS (MuoniuM lAser SpectroScopy) experiment}.
\newblock {\em Hyperfine Interact.}, 239(1):49, 2018.

\bibitem{eides2016hyperfine}
Michael~I Eides and Valery~A Shelyuto.
\newblock Hyperfine splitting in muonium and positronium.
\newblock {\em International Journal of Modern Physics A}, 31(28n29):1645034,
  2016.

\bibitem{eidesOsaka2018}
Michael~I Eides.
\newblock Hyperfine splitting in muonium: Theory meets experiment, talk
  presented at the international workshop on physics of muonium and related
  topics.
\newblock 2018.

\bibitem{Morel:2020dww}
L\'eo Morel, Zhibin Yao, Pierre Clad\'e, and Sa\"\i{}da Guellati-Kh\'elifa.
\newblock {Determination of the fine-structure constant with an accuracy of 81
  parts per trillion}.
\newblock {\em Nature}, 588(7836):61--65, 2020.

\bibitem{tanaka2021development}
K.~S. Tanaka, M.~Iwasaki, O.~Kamigaito, S.~Kanda, N.~Kawamura, Y.~Matsuda,
  T.~Mibe, S.~Nishimura, N.~Saito, N.~Sakamoto, S.~Seo, K.~Shimomura,
  P.~Strasser, K.~Suda, T.~Tanaka, H.~A. Torii, A.~Toyoda, Y.~Ueno, and
  M.~Yoshida.
\newblock Development of microwave cavities for measurement of muonium
  hyperfine structure at j-parc, 2021.

\bibitem{Grinin1061}
Alexey Grinin, Arthur Matveev, Dylan~C. Yost, Lothar Maisenbacher, Vitaly
  Wirthl, Randolf Pohl, Theodor~W. H{\"a}nsch, and Thomas Udem.
\newblock Two-photon frequency comb spectroscopy of atomic hydrogen.
\newblock {\em Science}, 370(6520):1061--1066, 2020.

\bibitem{Parker:2018vye}
Richard~H. Parker, Chenghui Yu, Weicheng Zhong, Brian Estey, and Holger
  M\"uller.
\newblock {Measurement of the fine-structure constant as a test of the Standard
  Model}.
\newblock {\em Science}, 360:191, 2018.

\bibitem{2020Natur.588...37M}
Holger {M{\"u}ller}.
\newblock {Standard model of particle physics tested by the fine-structure
  constant}.
\newblock {\em \nat}, 588(7836):37--38, January 2020.

\bibitem{Eides:2018rph}
Michael~I. Eides.
\newblock {Hyperfine Splitting in Muonium: Accuracy of the Theoretical
  Prediction}.
\newblock {\em Phys. Lett. B}, 795:113--116, 2019.

\bibitem{2016-+CODATA14}
Peter~J. Mohr, David~B. Newell, and Barry~N. Taylor.
\newblock Codata recommended values of the fundamental physical constants:
  2014.
\newblock {\em Rev. Mod. Phys.}, 88:035009, Sep 2016.

\bibitem{Mohr:2000ie}
P.~J. Mohr and B.~N. Taylor.
\newblock {CODATA Recommended Values of the Fundamental Physical Constants:
  1998}.
\newblock {\em Rev. Mod. Phys.}, 72:351--495, 2000.

\bibitem{Mariam:1982bq}
F.~G. Mariam et~al.
\newblock {HIGHER PRECISION MEASUREMENT OF THE HFS INTERVAL OF MUONIUM AND OF
  THE MUON MAGNETIC MOMENT}.
\newblock {\em Phys. Rev. Lett.}, 49:993--996, 1982.

\bibitem{Karshenboim:2021jsc}
Savely~G. Karshenboim and Valery~A. Shelyuto.
\newblock {Hadronic vacuum-polarization contribution to various QED
  observables}.
\newblock {\em Eur. Phys. J. D}, 75(2):49, 2021.

\bibitem{Eides:2014xea}
Michael~I. Eides and Valery~A. Shelyuto.
\newblock {Three-Loop Contributions to Hyperfine Splitting: Muon Loop
  Light-by-Light Insertion and Other Closed Lepton Loops}.
\newblock {\em Phys. Rev. D}, 90(11):113002, 2014.

\bibitem{burkley2021stable}
Zakary Burkley, Lucas de~Sousa~Borges, Ben Ohayon, Artem Golovozin, Jesse
  Zhang, and Paolo Crivelli.
\newblock Stable high power deep-uv enhancement cavity in ultra high vacuum
  with fluoride coatings, 2021.

\bibitem{PhysRevLett.72.2793}
E.~Morenzoni, F.~Kottmann, D.~Maden, B.~Matthias, M.~Meyberg, Th. Prokscha, Th.
  Wutzke, and U.~Zimmermann.
\newblock Generation of very slow polarized positive muons.
\newblock {\em Phys. Rev. Lett.}, 72:2793--2796, Apr 1994.

\bibitem{PhysRevLett.108.143401}
A.~Antognini, P.~Crivelli, T.~Prokscha, K.~S. Khaw, B.~Barbiellini, L.~Liszkay,
  K.~Kirch, K.~Kwuida, E.~Morenzoni, F.~M. Piegsa, Z.~Salman, and A.~Suter.
\newblock Muonium emission into vacuum from mesoporous thin films at cryogenic
  temperatures.
\newblock {\em Phys. Rev. Lett.}, 108:143401, Apr 2012.

\bibitem{Yerokhin:2018gna}
Vladimir~A. Yerokhin, Krzysztof Pachucki, and Vojtech Patkos.
\newblock {Theory of the Lamb Shift in Hydrogen and Light Hydrogen-Like Ions}.
\newblock {\em Annalen Phys.}, 531(5):1800324, 2019.

\bibitem{karr2020proton}
Jean-Philippe Karr, Dominique Marchand, and Eric Voutier.
\newblock The proton size.
\newblock {\em Nature Reviews Physics}, 2(11):601--614, 2020.

\bibitem{Pohl:2010zza}
Randolf Pohl et~al.
\newblock {The size of the proton}.
\newblock {\em Nature}, 466:213--216, 2010.

\bibitem{Antognini:1900ns}
Aldo Antognini et~al.
\newblock {Proton Structure from the Measurement of $2S-2P$ Transition
  Frequencies of Muonic Hydrogen}.
\newblock {\em Science}, 339:417--420, 2013.

\bibitem{PhysRevA.93.022513}
M.~Horbatsch and E.~A. Hessels.
\newblock Tabulation of the bound-state energies of atomic hydrogen.
\newblock {\em Phys. Rev. A}, 93:022513, Feb 2016.

\bibitem{torii2015precise}
Hiroyuki~A Torii, HA~Torii, M~Aoki, Y~Fukao, Y~Higashi, T~Higuchi, H~Iinuma,
  Y~Ikedo, K~Ishida, M~Iwasaki, et~al.
\newblock Precise measurement of muonium hfs at j-parc muse.
\newblock In {\em Proceedings of the 2nd International Symposium on Science at
  J-PARC—Unlocking the Mysteries of Life, Matter and the Universe—}, page
  025018, 2015.

\bibitem{Ueno:2018hdr}
Yasuhiro Ueno et~al.
\newblock {New Precision Measurement of Muonium Hyperfine Structure}.
\newblock {\em PoS}, ICHEP2018:466, 2019.

\bibitem{nishimura2021rabioscillation}
S.~Nishimura, H.~A. Torii, Y.~Fukao, T.~U. Ito, M.~Iwasaki, S.~Kanda,
  K.~Kawagoe, D.~Kawall, N.~Kawamura, N.~Kurosawa, Y.~Matsuda, T.~Mibe,
  Y.~Miyake, N.~Saito, K.~Sasaki, Y.~Sato, S.~Seo, P.~Strasser, T.~Suehara,
  K.~S. Tanaka, T.~Tanaka, J.~Tojo, A.~Toyoda, Y.~Ueno, T.~Yamanaka,
  T.~Yamazaki, H.~Yasuda, T.~Yoshioka, and K.~Shimomura.
\newblock Rabi-oscillation spectroscopy of the hyperfine structure of muonium
  atoms, 2021.

\bibitem{asaka2018precision}
T~Asaka, M~Tanaka, K~Tsumura, and M~Yoshimura.
\newblock Precision electroweak shift of muonium hyperfine splitting.
\newblock {\em arXiv preprint arXiv:1810.05429}, 2018.

\bibitem{1868388}
Alex Keshavarzi, Kim~Siang Khaw, and Tamaki Yoshioka.
\newblock {Muon $g-2$: current status}.
\newblock 6 2021.

\bibitem{Kanda:2020mmc}
S.~Kanda et~al.
\newblock {New precise spectroscopy of the hyperfine structure in muonium with
  a high-intensity pulsed muon beam}.
\newblock {\em Phys. Lett. B}, 815:136154, 2021.

\bibitem{kiselev2021progress}
D~Kiselev, P~Baumann, P~Duperrex, S~Jollet, P-R Kettle, A~Knecht, D~Laube,
  C~Nyfeler, A~Papa, D~Reggiani, et~al.
\newblock Progress and challenges of the psi meson targets and relevant
  systems.
\newblock In {\em Proceedings of the 3rd J-PARC Symposium (J-PARC2019)}, page
  011102, 2021.

\bibitem{kiselev2021status}
Daniela Kiselev, Christian Baumgarten, Rudolf D{\"o}lling, Pierre Duperrex,
  Dietmar G{\"o}tz, Joachim Grillenberger, Davide Reggiani, Markus Schneider,
  Marco Schippers, Mike Seidel, et~al.
\newblock Status and future projects of the psi high intensity proton
  accelerator.
\newblock In {\em Proceedings of the 3rd J-PARC Symposium (J-PARC2019)}, page
  011004, 2021.

\bibitem{knecht2017high}
A~Knecht.
\newblock The high intensity muon beam line (himb) project at psi.
\newblock In {\em The 19th International Workshop on Neutrinos from
  Accelerators (NUFACT2017). Uppsala}, 2017.

\bibitem{belosevic2019mucool}
Ivana Belosevic, Aldo Antognini, Y~Bao, Andreas Eggenberger, Malte Hildebrandt,
  Ryoto Iwai, DM~Kaplan, Kim~S Khaw, Klaus Kirch, Andreas Knecht, et~al.
\newblock mucool: a next step towards efficient muon beam compression.
\newblock {\em The European Physical Journal C}, 79(5):1--9, 2019.

\bibitem{PhysRevLett.125.164802}
A.~Antognini, N.~J. Ayres, I.~Belosevic, V.~Bondar, A.~Eggenberger,
  M.~Hildebrandt, R.~Iwai, D.~M. Kaplan, K.~S. Khaw, K.~Kirch, A.~Knecht,
  A.~Papa, C.~Petitjean, T.~J. Phillips, F.~M. Piegsa, N.~Ritjoho, A.~Stoykov,
  D.~Taqqu, and G.~Wichmann.
\newblock Demonstration of muon-beam transverse phase-space compression.
\newblock {\em Phys. Rev. Lett.}, 125:164802, Oct 2020.

\bibitem{antogninimucool}
A~Antognini and D~Taqqu.
\newblock mucool: muon cooling for high-brightness $\mu$ beams.
\newblock {\em SciPost Physics Proceedings}, 2021.

\bibitem{Keshavarzi:2021eqa}
Alex Keshavarzi, Kim~Siang Khaw, and Tamaki Yoshioka.
\newblock {Muon $g-2$: current status}.
\newblock 6 2021.

\bibitem{Eides:2021wuv}
Michael~I. Eides and Valery~A. Shelyuto.
\newblock {Three-Loop Spin-Independent Radiative-Recoil Corrections to Energy
  Levels in Muonium}.
\newblock 10 2021.

\bibitem{grinin2020two}
Alexey Grinin, Arthur Matveev, Dylan~C Yost, Lothar Maisenbacher, Vitaly
  Wirthl, Randolf Pohl, Theodor~W H{\"a}nsch, and Thomas Udem.
\newblock Two-photon frequency comb spectroscopy of atomic hydrogen.
\newblock {\em Science}, 370(6520):1061--1066, 2020.

\bibitem{Ubachs1033}
Wim Ubachs.
\newblock Crisis and catharsis in atomic physics.
\newblock {\em Science}, 370(6520):1033--1033, 2020.

\bibitem{Bezginov1007}
N.~Bezginov, T.~Valdez, M.~Horbatsch, A.~Marsman, A.~C. Vutha, and E.~A.
  Hessels.
\newblock A measurement of the atomic hydrogen lamb shift and the proton charge
  radius.
\newblock {\em Science}, 365(6457):1007--1012, 2019.

\bibitem{HAMMER2020257}
Hans-Werner Hammer and Ulf-G. Meißner.
\newblock The proton radius: from a puzzle to precision.
\newblock {\em Science Bulletin}, 65(4):257--258, 2020.

\bibitem{cui2021fresh}
Zhu-Fang Cui, Daniele Binosi, Craig~D. Roberts, and Sebastian~M. Schmidt.
\newblock Fresh extraction of the proton charge radius from electron
  scattering, 2021.

\bibitem{Pachucki:2018yxe}
Krzysztof Pachucki, Vojt\v{e}ch Patk\'o\v{s}, and Vladimir~A. Yerokhin.
\newblock {Three-photon exchange nuclear structure correction in hydrogenic
  systems}.
\newblock {\em Phys. Rev. A}, 97(6):062511, 2018.

\bibitem{Blokland:2001fn}
Ian~Richard Blokland, Andrzej Czarnecki, and Kirill Melnikov.
\newblock {Expansion of bound state energies in powers of m / M and (1-m / M)}.
\newblock {\em Phys. Rev. D}, 65:073015, 2002.

\bibitem{Beyer:2017gug}
Axel Beyer et~al.
\newblock {The Rydberg constant and proton size from atomic hydrogen}.
\newblock {\em Science}, 358(6359):79--85, 2017.

\bibitem{Liu:1995hq}
Weiwen Liu et~al.
\newblock {Observation of resonance line narrowing for old muonium}.
\newblock {\em Phys. Rev. A}, 52:1948--1953, 1995.

\bibitem{PhysRevA.2.1411}
B.~K. Rao, D.~Ikenberry, and T.~P. Das.
\newblock Hyperfine pressure shift and van der waals interaction. iv.
  hydrogen-rare-gas systems.
\newblock {\em Phys. Rev. A}, 2:1411--1421, Oct 1970.

\bibitem{Frugiuele:2019drl}
Claudia Frugiuele, Jes\'us P\'erez-R\'\i{}os, and Clara Peset.
\newblock {Current and future perspectives of positronium and muonium
  spectroscopy as dark sectors probe}.
\newblock {\em Phys. Rev. D}, 100(1):015010, 2019.

\bibitem{Karshenboim:2010cg}
S.~G. Karshenboim.
\newblock {Precision physics of simple atoms and constraints on a light boson
  with ultraweak coupling}.
\newblock {\em Phys. Rev. Lett.}, 104:220406, 2010.

\bibitem{Jackiw:1972jz}
R.~Jackiw and Steven Weinberg.
\newblock {Weak interaction corrections to the muon magnetic moment and to
  muonic atom energy levels}.
\newblock {\em Phys. Rev. D}, 5:2396--2398, 1972.

\bibitem{Jegerlehner:2009ry}
Fred Jegerlehner and Andreas Nyffeler.
\newblock {The Muon g-2}.
\newblock {\em Phys. Rept.}, 477:1--110, 2009.

\bibitem{Balkin:2021rvh}
Reuven Balkin, C\'edric Delaunay, Michael Geller, Enrique Kajomovitz, Gilad
  Perez, Yogev Shpilman, and Yotam Soreq.
\newblock {A Custodial Symmetry for Muon g-2}.
\newblock 4 2021.

\bibitem{messiah_2014}
Albert Messiah.
\newblock {\em Quantum mechanics}.
\newblock Dover Publications, 2014.

\bibitem{Salpeter:1951sz}
E.~E. Salpeter and H.~A. Bethe.
\newblock {A Relativistic equation for bound state problems}.
\newblock {\em Phys. Rev.}, 84:1232--1242, 1951.

\bibitem{Hanneke:2008tm}
D.~Hanneke, S.~Fogwell, and G.~Gabrielse.
\newblock {New Measurement of the Electron Magnetic Moment and the Fine
  Structure Constant}.
\newblock {\em Phys. Rev. Lett.}, 100:120801, 2008.

\bibitem{Hardy:2016kme}
Edward Hardy and Robert Lasenby.
\newblock {Stellar cooling bounds on new light particles: plasma mixing
  effects}.
\newblock {\em JHEP}, 02:033, 2017.

\bibitem{Fan:2020ijg}
X.~Fan and G.~Gabrielse.
\newblock {Driven One-Particle Quantum Cyclotron}.
\newblock {\em Phys. Rev. A}, 103(2):022824, 2021.

\bibitem{Gabrielse:2019cgf}
G.~Gabrielse, S.~E. Fayer, T.~G. Myers, and X.~Fan.
\newblock {Towards an Improved Test of the Standard Model\textquoteright{}s
  Most Precise Prediction}.
\newblock {\em Atoms}, 7(2):45, 2019.

\bibitem{Ilten:2018crw}
Philip Ilten, Yotam Soreq, Mike Williams, and Wei Xue.
\newblock {Serendipity in dark photon searches}.
\newblock {\em JHEP}, 06:004, 2018.

\bibitem{Andreev:2021xpu}
Yu.~M. Andreev et~al.
\newblock {Constraints on New Physics in Electron $g-2$ from a Search for
  Invisible Decays of a Scalar, Pseudoscalar, Vector, and Axial Vector}.
\newblock {\em Phys. Rev. Lett.}, 126(21):211802, 2021.

\bibitem{CODATANP}
C.~Delaunay et~al.
\newblock work in progress.

\bibitem{Abe:2019thb}
M.~Abe et~al.
\newblock {A New Approach for Measuring the Muon Anomalous Magnetic Moment and
  Electric Dipole Moment}.
\newblock {\em PTEP}, 2019(5):053C02, 2019.

\bibitem{Janka:2020xky}
G.~Janka et~al.
\newblock {Intense beam of metastable Muonium}.
\newblock {\em Eur. Phys. J. C}, 80(9):804, 2020.

\bibitem{ohayon2021precision}
B~Ohayon, G~Janka, I~Cortinovis, Z~Burkley, De~Bourges-Sousa, E~Depero,
  A~Golovizin, X~Ni, Z~Salman, A~Suter, et~al.
\newblock Precision measurement of the lamb shift in muonium.
\newblock {\em arXiv preprint arXiv:2108.12891}, 2021.

\end{thebibliography}

\clearpage
\newpage
\maketitle
\onecolumngrid

\begin{center}
\textbf{\large Towards an independent determination of muon g-2 from muonium spectroscopy} \\
\vspace{0.05in}
{ \it \large Supplemental Material}\\
\vspace{0.05in}
{C\'edric Delaunay, Ben Ohayon and Yotam Soreq}
\end{center}

\onecolumngrid
\setcounter{equation}{0}
\setcounter{figure}{0}
\setcounter{table}{0}
\setcounter{section}{0}
\setcounter{page}{1}
\makeatletter
\renewcommand{\theequation}{S\arabic{equation}}
\renewcommand{\thefigure}{S\arabic{figure}}
\renewcommand{\thetable}{S\arabic{table}}
\newcommand\ptwiddle[1]{\mathord{\mathop{#1}\limits^{\scriptscriptstyle(\sim)}}}


\section{Least-square adjustment}

The electron-to-muon mass ratio $m_e/m_\mu$ and the muon anomalous magnetic moment $a_\mu$ are extracted from muonium data through a least-square adjustment. We follow the method and notation used by the CODATA collaboration, as described in their 1998 review~\cite{Mohr:2000ie}. Our adjustment is based on the 1999~\cite{Meyer:1999cx}  measurement of the 1S-2S transition frequency at the Rutherford Appleton Laboratory (RAL), the 1982~\cite{Mariam:1982bq} and 1999~\cite{Liu:1999iz} measurements of the ground state HFS at the Clinton P. Anderson Meson Physics Facility (LAMPF). It includes also inputs $\delta E({\rm 1S})$, $\delta E({\rm 2S})$ and $\delta E({\rm HFS})$ representing additive corrections to the 1S, 2S energies $E_M$ and the 1S HFS, respectively, associated to unknown high-order QED contributions. They are assigned zero values and an uncertainty given by the theory uncertainty of the energy level calculations. Input data are summarized in Table~\ref{tab:inputs}. 

\begin{table}[h]
    \centering
    \begin{tabular}{c|c|c|c|c}
\hline\hline
input datum & value & relative uncertainty &  identification & reference \\
\hline
    $\nu_{\rm 1S-2S}$     & $2\,455\,528\,941.0(9.8)\,{\rm MHz}$ & $4.0\times 10^{-9}$ & RAL-99 &\cite{Meyer:1999cx}\\
    $\nu_{\rm HFS}$ & $4\,463\,302\,776(51)\,{\rm Hz} $& $1.2\times 10^{-8}$ & LAMPF-99 & \cite{Liu:1999iz}\\
    $\nu_{\rm HFS}$ & $4\,463\,302.88(16){\rm kHz}$ & $3.6\times10^{-8}$  &LAMPF-82 & \cite{Mariam:1982bq}\\
    $\delta E({\rm 1S})/h$ & $0.000(14)\,{\rm MHz}$ & $4.3\times 10^{-12}$ & theory & \cite{2018-CODATA} \\
    $\delta E({\rm 2S})/h$ & $0.0(1.8)\,{\rm kHz}$ & $2.2\times 10^{-12}$ & theory & \cite{2018-CODATA}\\
    $\delta E({\rm HFS})/h$ & $0.000(70)\,{\rm kHz}$ & $1.6\times 10^{-8}$ & theory & \cite{Eides:2018rph}
\\\hline\hline
    \end{tabular}
    \caption{Input data for the determination of $m_e/m_\mu$ and $a_\mu$ from muonium spectroscopy. For  the additive corrections $\delta E({\rm 1S})$ and $\delta E({\rm 2S})$ the quoted $u_r$ are relative to the corresponding state energies, and to the 1S HFS frequency for $\delta E({\rm HFS})/h$. }
    \label{tab:inputs}
\end{table}

The data is compared to the best available theory predictions for $\nu_{\rm 1S-2S}$ and $\nu_{\rm HFS}$. The full expressions in terms of $R_\infty$, $\alpha$, $m_e/m_\mu$ and $a_\mu$ (including latest developments in the $n$S levels energies~\cite{Yerokhin:2018gna} and the 1S HFS~\cite{Karshenboim:2021jsc,Eides:2014xea}) are nicely summarized in the 2018 CODATA review~\cite{2018-CODATA}. (The muonium energies are obtained from hydrogen ones upon changing the reduced mass and discarding the nuclear finite-size and polarizability contributions.) Given their very small uncertainty compared to those in muonium, we assume $R_\infty$ and $\alpha$ are exactly known constants fixed to their 2018 CODATA recommended values~\cite{2018-CODATA}: $R_\infty c=3\,289\,841\,960\,250.8\,$kHz and $\alpha^{-1}=137.035\,999\,084$. The observational equations relating the input data in Table~\ref{tab:inputs} and the remaining adjusted constants including $m_e/m_\mu$, $a_\mu$ and the additive corrections $\delta^{\rm th}_{\rm 1S}$, $\delta^{\rm th}_{\rm 2S}$, $\delta^{\rm th}_{\rm HFS}$ are given in Table~\ref{tab:obseqs}.

\begin{table}[h]
    \centering
    \begin{tabular}{c|c}
    \hline\hline
         input datum &  observational equation \\
         \hline
$\nu_{\rm 1S-2S}$      & $\nu_{\rm 1S-2S} \doteq [E_M({\rm 2S};m_e/m_\mu)+ \delta_{\rm 2S}^{\rm th}-E_M({\rm 1S};m_e/m_\mu) -\delta_{\rm 1S}^{\rm th}]/h$\\
$\nu_{\rm HFS}$ & $\nu_{\rm HFS} \doteq \nu_{\rm HFS}^{\rm th}(m_e/m_\mu,a_\mu)+\delta_{\rm HFS}^{\rm th}/h$\\
$\delta E({\rm 1S})/h$ & $\delta E({\rm 1S}) \doteq \delta_{\rm 1S}^{\rm th}$\\
$\delta E({\rm 2S})/h$ & $\delta E({\rm 2S}) \doteq \delta_{\rm 2S}^{\rm th}$\\
$\delta E({\rm HFS})/h$ & $\delta E({\rm HFS}) \doteq \delta_{\rm HFS}^{\rm th}$\\
\hline\hline
    \end{tabular}
    \caption{Observational equations used for the determination of $m_e/m_\mu$ and $a_\mu$ from muonium spectroscopy.}
    \label{tab:obseqs}
\end{table}

The theory uncertainty of the $n$S energy calculation is dominated by the unknown higher-order terms in the relativistic-recoil QED contribution. Reference~\cite{Meyer:1999cx} quotes a QED uncertainty of $20\,$kHz for the overall 1S-2S transition frequency. To our knowledge there is no updated value published in the literature. Following Ref.~\cite{2018-CODATA}, we reevaluated the muonium 1S and 2S energy uncertainties using the latest developments in hydrogen calculations, yielding a total uncertainty of $u[\delta E({\rm 1S})/h]\simeq14\,$kHz and $u[\delta E({\rm 2S})/h]\simeq1.8\,$kHz, with a $99\%$ correlation between them.

After two iterations, the least-square adjustment described above converged to the values reported in the main text: $m_e/m_\mu=4\,836\,329(4)\times 10^{-9}$ and $a_\mu=116\,637(82)\times 10^{-8}$, the former  reproducing the value derived in Ref.~\cite{Meyer:1999cx}. The Birge ratio $R_b=0.68$ and all normalized residuals are found to be below unity, showing self-consistency of the input data.  

The least-square adjustment presented above can be straightforwardly improved to include future more precise measurements and theory calculations (including other muonium transitions). Possible new physics contributions (as discussed below) can be accounted for by additional adjusted constants for the associated mass scale and couplings to electrons and muons.

\section{New physics contributions to muonium energy levels}

The new physics (NP) contributions from the exchange of a scalar or vector of mass $m$ is given in first-order perturbation theory by the overlap of the non-relativistic potential 
\begin{equation}\label{eq:VNP}
    V_{\rm NP}(r)= \alpha_{\rm NP}e^{-mr}\left[\frac{1}{r}+\frac{8\pi(\vec S_e\cdot \vec S_\mu)}{3m_em_\mu}\left[\frac{m^2}{4\pi r}-\delta^3(r)\right]\right]\,,
\end{equation}
with the modulus square of the Schr\"{o}dinger wavefunction for hydrogen-like atoms~\cite{messiah_2014}
\begin{equation}
    \psi_{nlm}(r,\theta,\phi)= R_{nl}(r)Y_l^m(\theta,\phi)\,,
\end{equation}
where $Y_l^m(\theta,\phi)$ are spherical harmonics (normalized as $\int_{-1}^1 d\cos\theta\int_0^{2\pi}d\phi\ Y_l^m(\theta,\phi)Y_{l'}^{m'}(\theta,\phi) = \delta_{ll'}\delta_{mm'}$) and 
\begin{equation}
    R_{nl}(r) = \sqrt{\left(\frac{2}{n a_M}\right)^3\frac{(n-l-1)!}{2n(n+l)!}}e^{-\rho/2}\rho^lL_{n-l+1}^{2l+1}(\rho)\,,\quad {\rm with}\quad \rho\equiv \frac{2r}{na_M}\,,
\end{equation}
is the radial wavefunction expressed in terms of the reduced Bohr radius $a_M=(1+m_e/m_\mu)/(\alpha m_e)$ and the generalized Laguerre polynomials of degree $n-l+1$, $L_{n-l+1}^{2l+1}(x)$.
The NP contribution to the 1S-2S frequency shift (arising from the spin-independent part of $V_{\rm NP}$) is
\begin{equation}
    \delta_{\rm 1S-2S}^{\rm NP} = \frac{8\alpha_{\rm NP}}{3\alpha} I_{12}(ma_M)\,,
\end{equation}
where
\begin{eqnarray}
    I_{12}(ma_M)&\equiv&a_M\int d^3r\, \frac{e^{-mr}}{r}\left[|\psi_{200}^2(r,\theta,\phi)|^2-|\psi_{100}^2(r,\theta,\phi)|^2\right]=a_M\int_0^\infty dr r e^{-mr}\left[R_{20}(r)^2-R_{10}(r)^2\right]\nonumber\\
    &=&\frac{1+2(ma_M)^2}{4(1+ma_M)^4}-\frac{4}{(2+ma_M)^2}\,,
\end{eqnarray}
with the asymptotic behaviors $I_{12}(x\to 0)=-3/4$ and $I_{12}(x\to \infty)\simeq -7/(2x^2)$.
Similarly, its counterpart to the 1S HFS shift (arising from the spin-spin interaction part of the potential) is 
\begin{equation}
    \delta_{\rm HFS}^{\rm NP}= \frac{\alpha_{\rm NP}}{4\alpha}J_1(m a_M)\,,
\end{equation}
where
\begin{eqnarray}
    J_1(ma_M) &\equiv& a_M^3\int d^3r\, e^{-mr}\left[\frac{m^2}{r}-4\pi\delta^3(r)\right]|\psi_{100}(r,\theta,\phi)|^2=a_M^3\left[-R_{10}(0)^2+m^2\int_0^\infty drre^{-mr}R_{10}(r)^2\right]\nonumber\\
    &=&-16\frac{1+ma_M}{(2+ma_M)^2}\,,
\end{eqnarray}
with $J_1(x\to 0)=-4$ and $J_1(x\to\infty)\simeq -16/x$. \\

Note that the large mass ($m\gg a_M^{-1}$) behavior of $J_1$ is at odds with naive expectations from effective field theory which predicts a decoupling of the HFS contribution like $1/m^2$ (or faster). The $1/m$ scaling is a spurious artefact of the non-relativistic approximation used to obtain the potential in Eq.~\eqref{eq:VNP}. While this approximation is certainly valid for long-range interactions, it breaks down for new spin-dependent forces whose range is shorter than the Compton wavelength of the electron. This is manifest from the appearance of a $\delta$-function term in Eq.~\eqref{eq:VNP} characterized by the scale $\sqrt{m_em_\mu}$.  This $\delta$-function stems from the assumption that the electron couples to the magnetic field created by a {\it static} muon located at $r=0$. For scalar/vector masses above that scale, the short-distance dynamics is not properly resolved, yielding a spurious $1/m$ scaling. The expected $1/m^2$-asymptotic of $J_1$ should be recovered within a bound-state theory, like the Bethe-Salpeter formalism~\cite{Salpeter:1951sz}, where both the electron {\it and} the antimuon are relativistic.

\begin{figure}[t]
    \centering
\begin{tabular}{cc}
    \includegraphics[width=0.47\columnwidth]{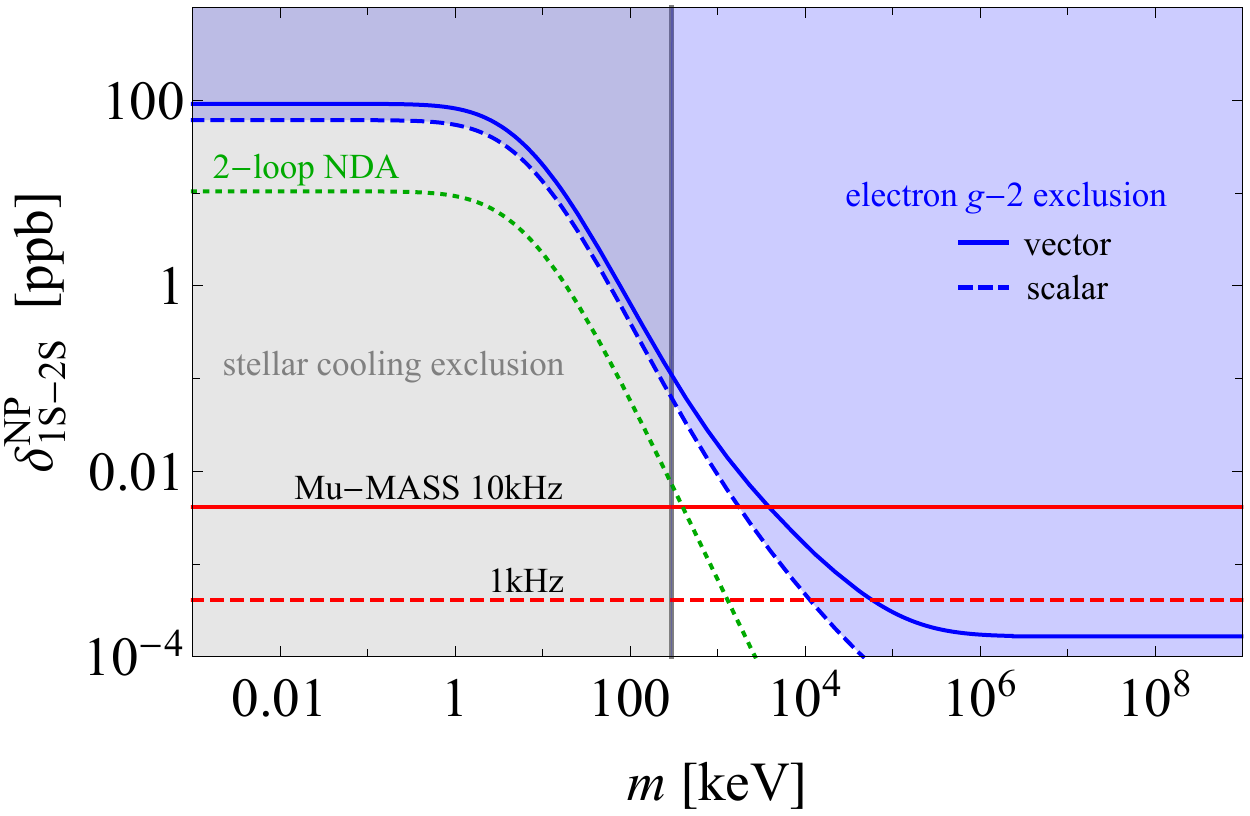}\hspace{0.5cm} &
    \includegraphics[width=0.47\columnwidth]{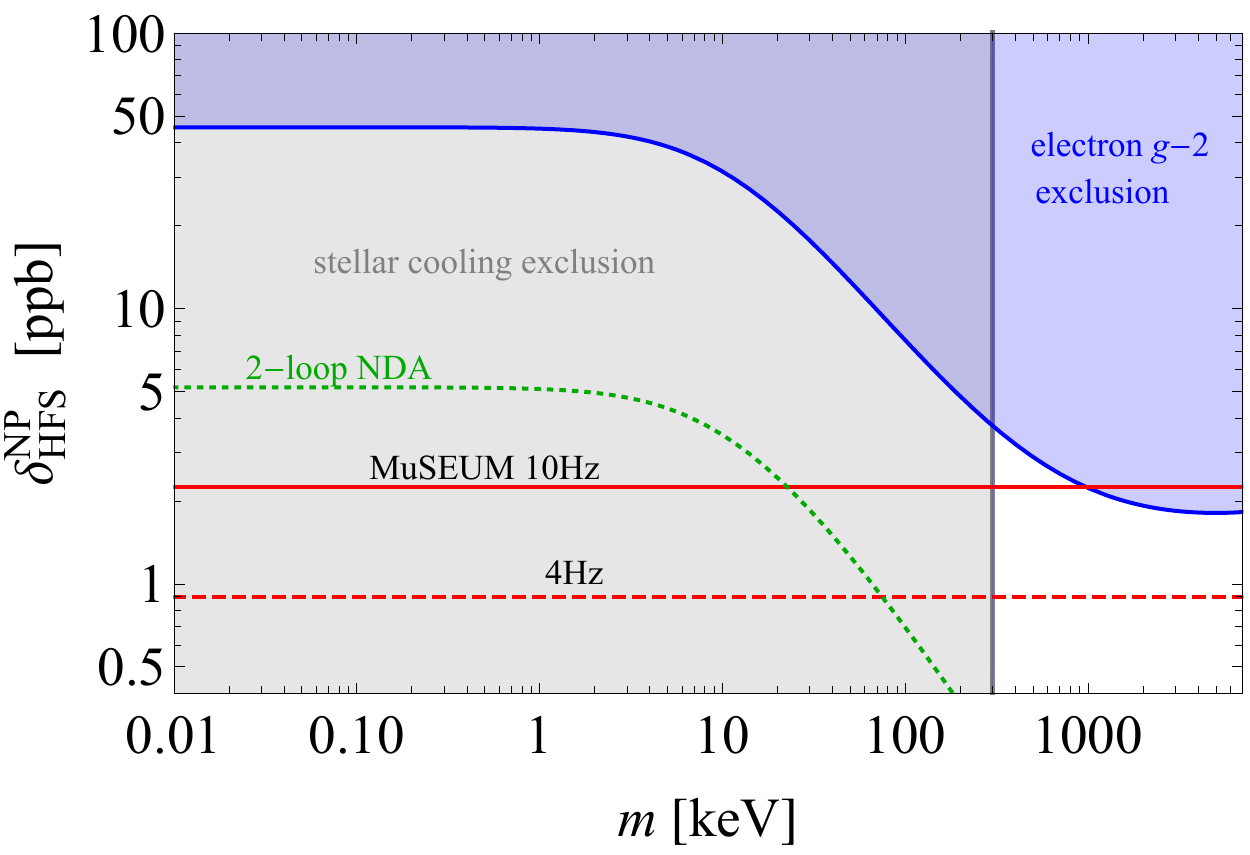}
\end{tabular}
    \caption{
    NP contributions to $\delta_{\rm 1S-2S}$ (left) and $\delta_{\rm HFS}$ (right). 
    Solid red lines indicate the expected accuracy of the ongoing Mu-Mass ($10\,$kHz) and MuSEUM ($10\,$Hz) experiments. 
    The blue (gray) shaded area is excluded by the electron $g-2$ measurement~\cite{Hanneke:2008tm} (stellar cooling constraints~\cite{Hardy:2016kme}). 
    The dotted green line denotes the NP contribution from the  relation $y_e=(\alpha/\pi)\times y_\mu$ naively expected from kinetic mixing at one-loop in the vector case. 
    The right-hand side plot is cut at $m=\sqrt{m_em_\mu}\simeq7\,$MeV, which is the validity range of the spin dependent effetive potentail in Eq.~\eqref{eq:VNP}. 
    }
    \label{fig:NP}
\end{figure}

To what extent NP could contaminate muonium observables given the discrepancy currently observed in muon $g-2$? In order to address this question, we fix the muon coupling $y_\mu$, assuming the NP contribution at one-loop to the anomalous magnetic moment of leptons~\cite{Jackiw:1972jz}
\beq
   \Delta a_{\ell}^{\rm NP}
=   \frac{y_\ell^2}{8\pi^2}
    \int_0^1dz\frac{(1-z)^2\zeta(z)}{(1-z)^2+z(m/m_\ell)^2}\,,
\eeq
[$\zeta(z)\equiv (1+z)$ for scalars and $2z$ for vectors] saturates Eq.~\eqref{eq:deltaamu}.
The electron counterpart is constrained by the good agreement between the measured electron $g-2$~\cite{Hanneke:2008tm} and the QED prediction~\cite{Aoyama:2019ryr} based on $\alpha$ determined from the recent $h/m_{\rm Rb}$ measurement at LKB~\cite{Morel:2020dww}, yielding ~\cite{Morel:2020dww}
\beq
    \Delta a_e
    \equiv 
    a_e^{\rm exp}-a_e^{\rm SM}(\alpha_{\rm LKB})
=   (4.8\pm3.0)\times 10^{-13}\,,    
\eeq
where the uncertainty is dominated by $a_e^{\rm exp}$. 
The maximal possible NP contributions (as function of $m$) to $\delta_{\rm 1S-2S}$ and $\delta_{\rm HFS}$, assuming $\Delta a_e^{\rm NP}\leq 10.8\times 10^{-13}$ and $\Delta a_\mu^{\rm NP}=259\times 10^{-11}$ are shown as blue lines in Fig.~\ref{fig:NP}. 
The NP contribution to the 1S-2S transition is suppressed above $a_M^{-1}$, due to the $1/m^2$ scaling of $\delta_{\rm 1S-2S}^{\rm NP}$, and goes below $10\,$kHz at about $m\sim 1\,$MeV. 
At lower masses, below $m\sim300\,$keV, the electron coupling is strongly constrained (gray lines) by stellar cooling  considerations~\cite{Hardy:2016kme}. The remaining gap can be covered by the order-of-magnitude improvement on $a_e^{\rm exp}$~\cite{Fan:2020ijg,Gabrielse:2019cgf}. 
This shows that the forthcoming 1S-2S frequency measurement at Mu-Mass will be typically safe from NP contamination, thus allowing a  determination $m_e/m_\mu$ at the ppb level. 
Similar conclusion holds for the $10\,$Hz accuracy goal on measuring the ground state HFS at MuSEUM. 
Note that, with additional assumptions on the scalar/vector decays, fixed-target experiments constrain the electron coupling more stringently up to one order of magnitude below the GeV-scale \textit{e.g.}~\cite{Ilten:2018crw,Andreev:2021xpu}. Taking these bounds at face value further excludes NP contributions to $\delta_{\rm 1S-2S}$ and $\delta_{\rm HFS}$ below the ultimate precision goal envisaged in the previous section.  

NP could also enter indirectly through a modification of the Rydberg constant (and in turn also $\alpha$). Since $R_\infty$ is determined from hydrogen spectroscopy, this is only possible assuming an additional coupling to protons. However, the overall consistency among several highly-precise hydrogen measurements strongly restricts NP interactions between electrons and protons, still allowing to extract $R_\infty$ with an uncertainty comparable to that of assuming the SM~\cite{CODATANP}. 
Hence we conclude that a reliable determination of $a_\mu$ from a comparison of future muonium spetroscopy measurements with their SM predictions is expected.  

\end{document}